\documentstyle[12pt,epsf]{article}

\advance\voffset by -1.5cm
\advance\hoffset by -1.5cm
\def\be{\begin{equation}}
\def\ee{\end{equation}}

\def\Zop{\bbbz}

\def\pmb#1{\setbox0=\hbox{#1}
 \kern-.025em\copy0\kern-\wd0
 \kern.05em\copy0\kern-\wd0
 \kern-.025em\raise.0433em\box0 }

\def\I{{\cal I}}

\def\3{\ss}
\def\sq{\hbox{\rlap{$\sqcap$}$\sqcup$}}
\def\qed{\ifmmode\sq\else{\unskip\nobreak\hfil
\penalty50\hskip1em\null\nobreak\hfil\sq
\parfillskip=0pt\finalhyphendemerits=0\endgraf}\fi}

\def\bbbz {{\sf Z\!\!Z}}

\newcommand{\ket}[1]{|#1\rangle}
\newcommand{\bra}[1]{\langle#1|}
\def\Tr{{\rm Tr}}

\def\ss{\bf S}

\def\N{{\cal N}}

\begin{document}

\thispagestyle{empty}
\def\thefootnote{\fnsymbol{footnote}}
\begin{flushright}
  hep-th/9806155\\
  HUTP-98-A050 \\
  DAMTP-1998-71
 \end{flushright}
\vskip 0.5cm

\begin{center}\LARGE
{\bf Stable non-BPS D-particles}
\end{center}
\vskip 1.0cm
\begin{center}
{\large  Oren Bergman\footnote{E-mail  address: 
{\tt bergman@string.harvard.edu}}}

\vskip 0.5 cm
{\it Lyman Laboratory of Physics\\
Harvard University\\
Cambridge, MA 02138}

\vskip 1.0 cm
{\large  Matthias R. Gaberdiel\footnote{E-mail  address: 
{\tt M.R.Gaberdiel@damtp.cam.ac.uk}}}

\vskip 0.5 cm
{\it Department of Applied Mathematics and Theoretical Physics \\
University of Cambridge, Silver Street, \\
Cambridge CB3 9EW, U.K.}
\end{center}

\vskip 1.0cm

\begin{center}
June 1998
\end{center}

\vskip 1.0cm

\begin{abstract}
It is shown that the orbifold of type IIB string theory by
$(-1)^{F_L}\,\I_4$ admits a stable non-BPS Dirichlet particle that is
stuck on the orbifold fixed plane. It is charged under the $SO(2)$
gauge group coming from the twisted sector, and transforms as a long
multiplet of the $D=6$ supersymmetry algebra. This suggests that it is
the strong coupling dual of the perturbative stable non-BPS state that
appears in the orientifold of type IIB by $\Omega\,\I_4$.
\end{abstract}

\vskip 1.0cm 
\begin{center}
PACS 11.25.-w, 11.25.Sq
\end{center}

\vfill
\setcounter{footnote}{0}
\def\thefootnote{\arabic{footnote}}
\newpage

\renewcommand{\theequation}{\thesection.\arabic
{equation}}

\section{Introduction}

Recently, it was observed by Sen \cite{Sen1} that duality symmetries
in string theory sometimes predict the existence of solitonic states
which are not BPS, but are stable due to the fact that they are the
lightest states carrying a given set of charge quantum numbers. The
most familiar example of this kind arises in the $Spin(32)/\Zop_2$
heterotic string, where the states in the spinor representation of the
gauge group are stable because of charge conservation, but are not BPS
as the right-movers are not in their ground states. Consequently, the
dual type I string theory should have a corresponding stable non-BPS
soliton. Other examples involve the open string state stretching
between a pair of Dirichlet $p$-branes of type II string theory on top
of an orientifold $p$-plane with $SO$ projection. In some of these
cases, (namely for $p=4$, $6$ and $7$), Sen gave an interpretation of
the corresponding dual solitonic state and determined its mass in the
appropriate limit \cite{Sen1}.

In this paper we shall consider the case corresponding to $p=5$ where
one of the theories in the dual pair is the orientifold of type IIB
string theory by $\Omega\,\I_4$. Under S-duality of type IIB, $\Omega$
is converted to $(-1)^{F_L}$ (where $F_L$ is the left-moving spacetime
fermion number), and the corresponding dual theory is the orbifold of
type IIB by $(-1)^{F_L}\I_4$. As we shall explain below, the spectrum
of the orbifold contains in the twisted sector a massless vector
multiplet of $\N=(1,1)$ supersymmetry in $D=6$, and this implies that
the orbifold fixed-plane corresponds to a (mirror) pair of D5-branes
on top of an orientifold 5-plane \cite{Sen1}. Because of the
orientifold projection, the massless states of the string stretching
between the two D5-branes is removed, and the gauge group is reduced
from $U(2)$ to $SO(2)$. The lightest state that is charged under the
$SO(2)$ is then the first excited open string state of the string
stretching between the two D5-branes: it forms a long multiplet of the
$\N=(1,1)$ $D=6$ supersymmetry, containing $128$ bosons and $128$
fermions. Since these states are stable, one should therefore expect
that the dual (orbifold) theory also contains a stable multiplet of
states that is charged under this $SO(2)$. It is the purpose of this
paper to construct the corresponding state (which shall turn out to be
an unconventional D-particle), and to show that it satisfies all the
required properties. A somewhat different proposal for this state has
recently been made by Sen \cite{Sen2}, who argued that it should be
described as a bound state of a D-string and an anti-D-string. It is
tempting to believe that the D-particle state we find corresponds
precisely to this bound state.
\medskip

The paper is organised as follows. In section~2 we shall review
briefly the boundary state approach to D-branes that we employ in the
following. In section~3 we describe in detail the spectrum of the
orbifold theory and construct the D-particle state.  We also
analyse its properties and show that it satisfies all the
requirements. We close with some open problems in section~4.

\section{Boundary states}
\setcounter{equation}{0}

Let us briefly review the boundary state approach to Dirichlet branes
\cite{CLNY,PolCai,GrGut,BG}. For simplicity we shall work in
light-cone gauge with $x^\pm = x^0 \pm x^9$ and transverse coordinates
$x^1, \ldots, x^8$. In this framework, a Dirichlet $p$-brane that is
parallel to the coordinate axes satisfies Neumann boundary conditions
for $x^i$ with $i=1,\ldots,p+1$, and Dirichlet boundary conditions for
$x^\pm$ and $x^I$ with $I=p+2,\ldots,8$.\footnote{Strictly speaking
these states are related by a double Wick rotation to Dirichlet branes
\cite{GrGut,BG}.} The corresponding boundary state is then of the form
\be
 \ket{Bp,\eta} = \exp\left\{
 \sum_{n>0}{1\over n}\Big[
  \alpha_{-n}^I\widetilde{\alpha}_{-n}^I
  -\alpha_{-n}^i\widetilde{\alpha}_{-n}^i
  \Big]
 + i\eta \sum_{r>0}\Big[
  \psi_{-r}^I\widetilde{\psi}_{-r}^I
  -\psi_{-r}^i\widetilde{\psi}_{-r}^i
  \Big]\right\}
 \ket{Bp,\eta}^{(0)} \,.
\label{b_state}
\ee
The parameter $\eta=\pm$ labels the different spin structures, and  
depending on the different sectors (untwisted or twisted, NS or R), 
the modings of the oscillators are either half-integral or integral. 
The ground state is usually taken to be an eigenstate of the momenta
in the Dirichlet directions for which bosonic zero modes exist
$\ket{Bp,\eta,k^\mu}^{(0)}$. In the untwisted sectors, $\mu$
runs over all $9-p$ Dirichlet directions, whereas in the
twisted sectors some of these directions may not have a zero mode. 
Position eigenstates are obtained by integrating over these momenta,
and in the untwisted sectors we have for example
\be
 \ket{Bp,\eta,x^I=x^\pm=0}^{(0)} = \int d^{9-p}k\,
    \ket{Bp,\eta,k^I,k^\pm}^{(0)} \,.
\ee
In the following, when we write $|Bp,\eta\rangle^{(0)}$, we
shall always refer to the position eigenstate with $x^I=x^\pm=0$.

A {\em physical} boundary state is invariant under the GSO-projection
and the various discrete symmetries, and is usually a linear
combination of states of the form (\ref{b_state}). In some
situations, the state should also be invariant under some of the
supersymmetry transformations. In order to identify a physical
boundary state with a D-brane, {\it i.e.} an object on which open
strings can end, one must further demand that the open string spectrum
resulting from the presence of this state is consistent with that of
the closed sector of the theory, since open strings that begin and end
on the D-brane can close to give a closed string state.  The relevant
open string spectrum can be determined by computing a tree-level
two-point function (cylinder) of the boundary state (with itself), and
expressing the result as a trace over open string states (annulus),
\be
\label{closed_open}
 \int dl\bra{Bp,\eta}e^{-lH_c}\ket{Bp,\eta'} = 
 2V \int {dt\over 2t}\Tr_{open}e^{-2tH_o} \,,
\ee
where $V$ is the (infinite) volume of the Neumann directions,
and the factor of $2$ is due to the fact that there are two
orientations for the open string. The open string sectors that appear
on the right hand side of (\ref{closed_open})
depend both on the closed string sector and on the spin 
structures 
$\eta,\eta'$. The relation is summarised in table~1.
\begin{table}[hbt]
\begin{center}
\begin{tabular}{|c|c|c|l|} \hline
 closed string & $\eta$ & $\eta'$ & open string \\ \hline
 NSNS & $\pm$ & $\pm$ & NS \\
 NSNS & $\pm$ & $\mp$ & R \\
 RR & $\pm$ & $\pm$ & NS$(-1)^F$ \\ 
 RR & $\pm$ & $\mp$ & R$(-1)^F$ \\\hline
\end{tabular}
\caption{Closed and open string channels of the cylinder.}
\end{center}
\end{table}
\smallskip

\noindent As an example, consider type IIA/B string theory. 
The GSO projection is given by
\be
 P_{GSO} = {1\over 4}\left\{
 \begin{array}{rl}
  \Big(1+(-1)^{\widetilde{F}}\Big)
         \Big(1+ (-1)^{F}\Big) &
     \quad\mbox{NSNS} \\
  \Big(1\mp (-1)^{\widetilde{F}}\Big)
         \Big(1+ (-1)^{F}\Big) &
     \quad\mbox{RR}\; ,
 \end{array}\right.
\label{II_GSO}
\ee
where the tilde denotes left-movers. The NSNS sector has a
non-degenerate ground state 
$\ket{Bp,\eta}^{(0)}_{NSNS} =\ket{0}_{NSNS}$,  which is
taken to be {\em odd} under both $(-1)^F$ and
$(-1)^{\widetilde{F}}$. The bosonic oscillators 
$\alpha^\mu_n,\widetilde{\alpha}^\mu_n$ are integrally moded,
$n\in\bbbz$, and the fermionic oscillators
$\psi^\mu_r,\widetilde{\psi}^\mu_r$ are half-integrally moded,  
$r\in \bbbz+1/2$. The action of  $(-1)^F$ and $(-1)^{\widetilde{F}}$
on the boundary states is given by 
\be
 (-1)^F\ket{Bp,\eta}_{NSNS} = 
 (-1)^{\widetilde{F}}\ket{Bp,\eta}_{NSNS} =
 -\ket{Bp,-\eta}_{NSNS} \,,
\ee
and therefore the combination 
$(\ket{Bp,+}_{NSNS}-\ket{Bp,-}_{NSNS})$ 
is GSO invariant and corresponds to a physical boundary state. In the
RR sector the fermionic oscillators are integrally moded, 
$r\in\bbbz$,
and there are therefore sixteen fermionic zero modes given by
$\psi_0^{\mu}, \widetilde{\psi}_0^{\mu}$ with $\mu=1,\ldots,8$. If we
introduce  
\be
 \psi_{\pm}^\mu = {1\over\sqrt{2}}(\psi_0^\mu
  \pm i\widetilde{\psi}_0^\mu) \,,
\ee
the ground state in the RR sector corresponding to the spin-structure
$\eta$, $\ket{Bp,\eta}^{(0)}_{RR}$, satisfies\footnote{These
conventions are opposite to the ones in \cite{BG}, but agree with
\cite{Sen1}.} 
\begin{eqnarray}
 \psi_{\eta}^i\ket{Bp,\eta}^{(0)}_{RR} &=& 0\nonumber \\
 \psi_{-\eta}^I\ket{Bp,\eta}^{(0)}_{RR} &=& 0 \,.
\end{eqnarray}
We choose the relative normalisation between the states corresponding
to $\eta=\pm$ by defining
\be
 \ket{Bp,+}^{(0)}_{RR} = \prod_{i=1}^{p+1}\psi_+^i
  \prod_{I=p+2}^8\psi_-^I \ket{Bp,-}^{(0)}_{RR} \,.
\ee
It then follows that 
\be
\ket{Bp,-}^{(0)}_{RR} = \prod_{i=1}^{p+1}\psi_-^i
  \prod_{I=p+2}^8\psi_+^I \ket{Bp,+}^{(0)}_{RR} \,.
\ee
As in \cite{Sen2}, the GSO operators $(-1)^F$ and
$(-1)^{\widetilde{F}}$ act on the RR ground states as 
\be
 (-1)^F = \prod_{\mu=1}^{8} (\sqrt{2} \psi_0^\mu) \,,
 \qquad
 (-1)^{\widetilde{F}} =  
   \prod_{\mu=1}^{8} (\sqrt{2} \widetilde{\psi}_0^\mu) \,,
\ee
and consequently,
\begin{eqnarray}
\label{RRu}
 (-1)^F\ket{Bp,\eta}_{RR} &=& 
           \ket{Bp,-\eta}_{RR}\nonumber \\
 (-1)^{\widetilde{F}}\ket{Bp,\eta}_{RR} &=& 
           (-1)^{p+1} \ket{Bp,-\eta}_{RR} \,.
\end{eqnarray}
We therefore find that the combinations
$(\ket{Bp,+}_{RR}+\ket{Bp,-}_{RR})$ are physical provided 
that $p$ is 
even in IIA, and $p$ is odd in IIB. By combining NSNS and RR
boundary states we get states which preserve $1/2$ of the
supersymmetry, {\it i.e.} BPS states. These are the D-branes
(or anti-D-branes)
\be
 \ket{Dp} = (\ket{Bp,+}_{NSNS}-\ket{Bp,-}_{NSNS})
   \pm (\ket{Bp,+}_{RR}+\ket{Bp,-}_{RR})\,,
\ee
where the relative sign between the NSNS and the RR component
distinguishes branes from anti-branes.  Due to the above restriction
on $p$ in the RR component, there are only even $p$ D-branes in type
IIA, and only odd $p$ D-branes in IIB. The necessity of combining the
NSNS and RR boundary states can also be understood from the
open-closed consistency condition (see table~1): in order to get the
GSO-projected NS open string spectrum (without a tachyon), we need
both a contribution from NSNS and RR. Since the NSNS and RR components
are linear combinations of $\eta=\pm$, the entire open string spectrum
consists of NS$(1+(-1)^F)/2$ and R$(1+(-1)^F)/2$, which is indeed
supersymmetric.

\section{A D-particle in type IIB/$(-1)^{F_L}\I_4$}
\setcounter{equation}{0}

Let us now consider the orbifold of type IIB theory on
$R^{9,1}/(-1)^{F_L}\I_4$, where $F_L$ is the left-moving 
space-time
fermion number, and $\I_4$ denotes inversion of four spatial
coordinates, $x^5,\ldots,x^8$, say. The fixed points under 
$\I_4$ form
a 5-plane at $x^5=x^6=x^7=x^8=0$, which extends along the 
coordinates
$x^1,\ldots,x^4$, as well as the light-cone coordinates 
$x^0,x^9$.  
In light-cone gauge, type IIB string theory has 16 dynamical
supersymmetries and 16 kinematical supersymmetries.
The former transform under the transverse $SO(8)$ as
\be
 Q \sim {\bf{8}}_s\;, \quad 
  \widetilde{Q}\sim {\bf{8}}_s \,.
\ee 
The orbifold breaks the transverse $SO(8)$ into 
$SO(4)_S \times SO(4)_R$, 
where the $SO(4)_S$ factor corresponds to rotations of 
$(x^1,\ldots,x^4)$, and the $SO(4)_R$ factor to
rotations of $(x^5\ldots,x^8)$.
The above supercharges therefore decompose as 
\be
 \bf{8}_s \longrightarrow  
  ((\bf{2},\bf{1}),(\bf{2},\bf{1})) +
  ((\bf{1},\bf{2}),(\bf{1},\bf{2})) \,.
\ee
The operator $\I_4$ reverses the sign of the vector representation of
$SO(4)_R$ (the $(\bf{2},\bf{2})$), and we therefore choose its action
on the $SO(4)_R$ spinors as  
\be
 \I_4 : \quad \left\{
 \begin{array}{rcr}
  (\bf{2},\bf{1}) & \rightarrow & (\bf{2},\bf{1}) \\
  (\bf{1},\bf{2}) & \rightarrow & - (\bf{1},\bf{2}) 
 \end{array}\right.\;.
\ee
The action of $(-1)^{F_L}$ is simply
\be
 (-1)^{F_L} : \quad Q \rightarrow Q\;, \quad
   \widetilde{Q} \rightarrow - \widetilde{Q} \,,
\ee
and the surviving supersymmetries thus transform as 
\be
 Q \sim (({\bf{2}},{\bf{1}}),({\bf{2}},{\bf{1}}))\;,
 \quad
 \widetilde{Q} \sim 
    ((\bf{1},\bf{2}),(\bf{1},\bf{2})) \,.
\ee
{}From the point of view of the 5-plane world-volume this is
(dynamical, light-cone) $\N=(1,1)$ supersymmetry\footnote{The same
orbifold of type IIA would yield $\N=(2,0)$ supersymmetry.}.
\smallskip
 
The closed string spectrum consists of an untwisted sector containing 
type IIB states which are even under $(-1)^{F_L}\I_4$, and a
twisted sector which is localised at the 5-plane. In the twisted
sector the various oscillators are moded as 
\begin{eqnarray}
 \mbox{twisted NS} : & 
  n\in  \left\{
  \begin{array}{ll}
   \bbbz & \mu=1,\ldots,4 \\
   \bbbz + 1/2 & \mu = 5,\ldots,8 
  \end{array}\right.
  \qquad 
  r\in \left\{
  \begin{array}{ll}
   \bbbz+1/2 & \mu=1,\ldots,4 \\
   \bbbz & \mu = 5,\ldots,8 
  \end{array}\right. \nonumber \\
 \mbox{twisted R} : & 
  n\in  \left\{
  \begin{array}{ll}
   \bbbz & \mu=1,\ldots,4 \\
   \bbbz +1/2 & \mu = 5,\ldots,8 
  \end{array}\right.
  \qquad 
  r\in\left\{
  \begin{array}{ll}
   \bbbz & \mu=1,\ldots,4 \\
   \bbbz + 1/2 & \mu = 5,\ldots,8 \,.
  \end{array}\right. 
\label{twisted_moding}
\end{eqnarray}
The ground state energy vanishes in both the R and NS 
sectors, and they both contain four fermionic zero modes
that transform in the vector representation of 
$SO(4)_S$ and $SO(4)_R$, respectively.
Consequently the twisted NSNS and RR ground states transform as  
\be
\left( ({\bf 2}, {\bf 1}) +
       ({\bf 1}, {\bf 2}) \right) \otimes
\left( ({\bf 2}, {\bf 1}) + 
       ({\bf 1}, {\bf 2}) \right) \,,
\ee
where the charges correspond to $SO(4)_S$ ($SO(4)_R$)
in the RR (NSNS) sector.
The unique massless  representation of $D=6$ $\N=(1,1)$
supersymmetry (other than the gravity multiplet) is the 
vector multiplet 
\be
\label{vector_multiplet}
 (({\bf{2}},{\bf{2}}),({\bf{1}},{\bf{1}})) + 
 (({\bf{1}},{\bf{1}}),({\bf{2}},{\bf{2}})) +
 \mbox{fermions} \,.
\ee
In order to preserve supersymmetry, we therefore have to 
choose the GSO-projections in all twisted sectors to be of the 
form 
\be
\label{twisted_GSO}
 P_{GSO,T} = {1\over 4}\Big(1-(-1)^{\widetilde{F}}\Big)
             \Big(1+(-1)^{F}\Big) \,.
\ee
This agrees with what we would have expected from standard 
orbifold techniques, namely that the effect of $(-1)^{F_L}$ is 
to change the left-GSO projection in the twisted sector.
In addition, the spectrum of the twisted sector must be
projected onto a subspace with either $(-1)^{F_L}\I_4=+1$
or $(-1)^{F_L}\I_4=-1$ (in the untwisted sector only $+1$
is allowed). 
Since twisted NSNS (RR) states are even (odd) 
under $(-1)^{F_L}$, and $\I_4$ reverses the sign of the vector of
$SO(4)_R$ (and leaves the vector of $SO(4)_S$ invariant), 
we conclude that in the present case the twisted sector states are odd
under $(-1)^{F_L}\I_4$.
\bigskip

Having described the spectrum and the GSO projections of the 
various sectors in some detail, we can now analyse whether a
D-particle boundary state is permitted. In the (untwisted) NSNS sector
$(-1)^{F_L}$ acts trivially, and all boundary states of the form
(\ref{b_state}) are invariant under $\I_4$, since $\I_4$ acts in the
same way on left- and right-movers. We therefore have a physical 
$p=0$ NSNS boundary state   
\be
|U0\rangle = \left( |B0,+\rangle_{NSNS} - 
  |B0,-\rangle_{NSNS} \right) \,.
\ee
On the other hand the $p=0$ RR boundary state is not 
physical because of (\ref{RRu}).\footnote{This boundary 
state is also not invariant under $(-1)^{F_L}\I_4$, as 
follows from the analysis of \cite{Sen2}.}
In the twisted sector, the boundary state is of the same 
form (with the appropriate modings). Since there are only bosonic 
zero modes for $\mu=0,1,2,3,4,9$, and since $x^1$ is
a Neumann direction, the momentum integral is over the 5-dimensional
space corresponding to $\mu=0,2,3,4,9$. The ground states satisfy  
\be
 \psi^\nu_{\mp} 
   |B0,\pm\rangle_{NSNS,T}^{(0)} = 0
\qquad \mbox{for $\nu=5,6,7,8$,} 
\ee
in the twisted NSNS sector, and
\be
\begin{array}{lcll}
\psi^1_{\pm} |B0,\pm\rangle_{RR,T}^{(0)} & = & 0 &  \\[5pt]
\psi^\mu_{\mp} |B0,\pm\rangle_{RR,T}^{(0)} & = & 0 &
\mbox{for $\mu=2,3,4$,}
\end{array}
\ee
in the twisted RR sector. 
On the ground states, the GSO operators act as 
\be
\begin{array}{rcc}
 \mbox{twisted NSNS :}\quad & 
  (-1)^F = \prod_{\mu=5}^{8}(\sqrt{2} \psi_0^{\mu})\,, &
  (-1)^{\widetilde{F}} = \prod_{\mu=5}^{8} 
      (\sqrt{2} \widetilde{\psi}_0^{\mu}) \\[10pt]
 \mbox{twisted RR :}\quad &
  (-1)^F = \prod_{\mu=1}^{4} 
      (\sqrt{2} \psi_0^{\mu}) \,,&
  (-1)^{\widetilde{F}} = \prod_{\mu=1}^{4} 
      (\sqrt{2} \widetilde{\psi}_0^{\mu}) \,.
\end{array}
\ee
Using the same arguments as before in the untwisted sector we 
find  
\be
(-1)^F |B0,\pm\rangle_{NSNS,T} = |B0,\mp\rangle_{NSNS,T}\,, \quad
(-1)^{\widetilde{F}} |B0,\pm\rangle_{NSNS,T} = 
   + |B0,\mp\rangle_{NSNS,T} 
\ee
and
\be
(-1)^F |B0,\pm\rangle_{RR,T} = |B0,\mp\rangle_{RR,T}\,, \quad
(-1)^{\widetilde{F}} |B0,\pm\rangle_{RR,T} = 
  - |B0,\mp\rangle_{RR,T} \,. 
\ee
Because of (\ref{twisted_GSO}) it then follows that only the 
combination $( |B0,+\rangle_{RR,T} + |B0,-\rangle_{RR,T} )$
in the twisted RR sector survives the GSO-projection, and that 
no combination of twisted NSNS sector boundary states is GSO
invariant. In addition, the ground states of the twisted RR 
sector boundary state
are odd under $(-1)^{F_L}\I_4$, as they are precisely the
vector states of $SO(4)_S$ that arise in the twisted sector. 
We therefore have one further physical boundary state 
\be
|T0\rangle = \left( |B0,+\rangle_{RR,T} + 
  |B0,-\rangle_{RR,T} \right)\,,
\ee
and the total D-particle state is of the form
\be
 \ket{D0} = \N_{U}\ket{U0} + \N_{T}\ket{T0}\,.
\label{D-particle}
\ee
We can then determine the cylinder diagram for a closed string that
begins and ends on the D-particle, and we find that 
\be
 \int_0^\infty dl \bra{D0} e^{-lH_c} \ket{D0} = 
 \int_0^\infty {dt\over t^{3/2}}
  \left\{ 2^{9/2}\N_{U}^2
   {f_3^8(e^{-\pi t}) - f_2^8(e^{-\pi t})\over
         f_1^8(e^{-\pi t})}
 + 2^{5/2} \N_{T}^2 
    {f_3^4(e^{-\pi t})
        f_4^4(e^{-\pi t}) \over
          f_1^4(e^{-\pi t})f_2^4(e^{-\pi t})}\right\}\,,
\ee
where $f_i$ are the standard $f$-functions \cite{PolCai,BG}. 
For $\N_T^2 = 2^4 \N_U^2 = L/(\pi 2^3)$ we then obtain (compare
\cite{Sen2}) 
\be
 \int dl \bra{D0} e^{-lH_c} \ket{D0} = 
  2L \int{dt\over 2t} \Tr_{NS-R}
   \Big[{1\over 2}(1+(-1)^F\I_4)e^{-2tH_o}\Big] \,,
\ee
where $L$ is the (infinite) size of the $x^0$ direction.
The open string spectrum thus consists of NS
and R sectors, both projected by $1/2(1+(-1)^F\I_4)$.
The tachyon of the NS sector is even under $\I_4$ but
odd under $(-1)^F$, and is therefore removed from the 
spectrum. This indicates that the D-particle is stable.

In addition, 4 massless states are removed from the NS sector, 
leaving 4 massless bosons, and the R sector contains 8 massless 
fermions. Including the zero modes in the light-cone 
directions,\footnote{When counting the zero modes of a D-brane
one must include the light-cone directions as well as 
the physical (transverse) massless states of the open
string. See for example \cite{Witten_bound} for a discussion
of the type IIB D-string.} this gives the D-particle 5 bosonic zero
modes and 16 fermionic zero modes. The former reflect the fact that
the D-particle is restricted to moving within the 5-plane, and the
latter give rise to a long ($2^8=256$-dimensional) representation of
the six-dimensional $\N=(1,1)$ supersymmetry. Finally, the D-particle
is charged under the vector field in the twisted RR sector. We have
therefore managed to construct a boundary state that possesses all the
properties that we expected to find from the S-dual description. 
\medskip

It seems quite plausible that this D-particle is the elusive ground
state of the D-string anti-D-string system sought by Sen
\cite{Sen2}. In particular, the analysis of Sen suggests that the
bound state should correspond to a superposition of an untwisted NSNS
state and a twisted RR state, but should not have any other
components, and this is indeed what we find.\footnote{If we take the
superpositions of the states in section 3 of \cite{Sen2} literally,
then the other components cancel.} On the other hand, the mass of our
D-particle does not seem to agree with the mass predicted by Sen: the
D-particle has the same mass as a conventional D-particle in type IIA,
$M_{D0} = 1/(\sqrt{\alpha'} g_s)$, and this differs by a factor of
$\sqrt{2}$ from the formula given in \cite{Sen2}.

\section{Conclusions}
\setcounter{equation}{0}

In this paper we have constructed a Dirichlet particle boundary state
in the orbifold of type IIB by $(-1)^{F_L} \I_4$.  The D-particle is
stuck on the orbifold plane, and satisfies the properties that are
expected from the S-dual description. In particular, it is stable (as
the open string beginning and ending on the D-particle is
tachyon-free), it gives rise to a long multiplet of the
six-dimensional $\N=(1,1)$ supersymmetry, and it is charged under the
relevant $SO(2)$ gauge group.

On the other hand some open problems remain. In particular, 
since the open string that begins and ends on the D-particle has the
projection $1/2(1+(-1)^F \I_4)$, rather than $1/4(1+(-1)^F)(1+ \I_4)$,
the open string spectrum admits states that are $\I_4$ odd as well as
states that are $\I_4$ even. It is therefore not obvious that such a
spectrum is consistent with the closed string spectrum, given that
open and closed strings couple. On the other hand, it is quite
plausible that the open-closed consistency condition is indeed
satisfied, since the closed string spectrum also contains 
$\I_4$ even and $\I_4$ odd states. (Indeed, it is not clear either
whether the other projection would in fact give rise to a consistent
open-closed spectrum.) It would be interesting to check this in
detail.   
\smallskip

It would also be interesting to see whether a similar analysis can be
given for the case of the spinorial representation of the
$Spin(32)/\Zop_2$ heterotic string, whose corresponding dual should be
a non-BPS particle in type I.  It is again possible to have the NSNS
component of the boundary state, but in this case it is not clear what
should play the role of the twisted RR component, or how to get rid of
the open string tachyon.  The situation is therefore much less clear.

\section*{Acknowledgements}

We thank Ashoke Sen for very useful correspondence. 
We also thank Eric Gimon and Michael Green for helpful
conversations. 
O.B. is supported in part by the NSF under grant PHY-92-18167.
M.R.G. is grateful to Jesus College, Cambridge for a Research
Fellowship.


\begin{thebibliography}{[20]}

\bibitem{Sen1} A. Sen, {\it Stable non-BPS states in string theory}, 
{\sf hep-th/9803194}.

\bibitem{Sen2} A. Sen, {\it Stable non-BPS bound states of BPS
D-branes}, {\sf hep-th/9805019}.

\bibitem{CLNY} C.G. Callan, C. Lovelace, C.R. Nappi, S.A. Yost,
{\it Loop corrections to superstring equations of motion},
Nucl. Phys. {\bf B 308}, 221 (1988).

\bibitem{PolCai} J. Polchinski, Y. Cai, {\it Consistency of open
superstring theories}, Nucl. Phys. {\bf B 296}, 91 (1988).

\bibitem{GrGut} M.B. Green, M. Gutperle, {\it Light-cone
supersymmetry and $D$-branes}, Nucl. Phys. {\bf B 476}, 484
(1996); {\sf hep-th/9604091}.

\bibitem{BG} O. Bergman, M.R. Gaberdiel, {\it A non-supersymmetric
open string theory and S-duality}, Nucl. Phys.~{\bf B 499}, 183 (1997); 
{\sf hep-th/9701137}. 

\bibitem{Witten_bound} E. Witten, {\it Bound states of strings and
p-branes}, Nucl. Phys.~{\bf B 460}, 335 (1996); 
{\sf hep-th/9510135}.

\end{thebibliography}
\end{document}